\documentclass[aps, prl, 10pt, twocolumn, superscriptaddress, noshowpacs, preprintnumbers, floatfix]{revtex4-1}

\usepackage[dvipsnames]{xcolor}
\usepackage{mathtools}
\usepackage{amsmath}
\usepackage{amsfonts}
\usepackage{amssymb}
\usepackage{bbold}
\usepackage{upgreek}
\usepackage{multirow}
\usepackage{bm}
\long\def\exclude#1{}
\usepackage{orcidlink}
\usepackage{slashed}
\usepackage{url}

\usepackage{graphicx}
\usepackage{color}

\begin{document}

\title{Resonant Annihilation of WIMP Dark Matter for Halo Gamma Ray Signal}
% \preprint{}

\author{Hitoshi Murayama}
\affiliation{Leinweber Institute for Theoretical Physics, University of California, Berkeley, CA 94720, USA}
\affiliation{Kavli Institute for the Physics and Mathematics of the Universe (WPI), University of Tokyo, Kashiwa 277-8583, Japan}
\affiliation{Ernest Orlando Lawrence Berkeley National Laboratory, Berkeley, CA 94720, USA}

\begin{abstract}
We propose a resonant annihilation as a way to reconcile the WIMP annihilation cross sections in a recently reported gamma ray signal from the Milky Way halo with that for the freeze-out and the upper limit from dwarf galaxies. We perform a simple model-independent analysis based on this hypothesis and determine the required parameters. We also present a simple particle-physics model that can accommodate them.
\end{abstract}

\maketitle
%~\newpage

\section{Introduction}

Dark matter of the universe has been one of the biggest mysteries in physics  (see, {\it e.g.}\/, \cite{Cirelli:2024ssz} for a recent review). In 1930s, it was hypothesized by Fred Zwicky because the motion of galaxies in the coma cluster was much faster than what would have been possible with the gravity among galaxies alone. There were also preceding suggestions by Knut Landmark and Jan Oort. Vera Rubin and Kent Ford studied rotation curves of stars in galaxies and found evidence for dark matter in 1970s. Dark Matter became a crucial element in the structure formation theory by James Peebles and others. CMB observations and gravitational lensing made the evidence unmistakeable. Compared to the mass density of ordinary matter (electrons, protons, neutrons), dark matter comprises more than 5 times mass density on average in the universe today.

Weakly Interacting Massive Particle (WIMP) has long been the favorite candidate among particle physicists (see, {\it e.g.}\/, \cite{Murayama:2007ek} for a review). It is a thermal relic of a stable unknown particle produced in the early universe, whose abundance was drastically depleted by annihilation into the particles in the standard model. To obtain the correct abundance, the annihilation cross section at the freeze-out in the early universe needs to be \cite{Cirelli:2024ssz}
\begin{align}
	\langle \sigma_{ann} v_{rel} \rangle_{\rm f.o.}
	\approx 2.2 \times 10^{-26}~{\rm cm}^3/{\rm s}. \label{eq:sigmafo}
\end{align}
This size of the annihilation cross section is naturally obtained for particles in the TeV energy scale, where many theories on physics beyond the standard model predicted new particles. Yet WIMPs have so far defied searches at colliders such as the LHC or direct detection in underground experiments. On the other hand, there is an excess in gamma rays from the galactic center which may be interpreted as produced by annihilation of dark matter \cite{Goodenough:2009gk,Fermi-LAT:2017opo,Murgia:2020dzu}. Unfortunately, the galactic center has a high density of stars and unresolved pulsars may produce gamma rays, and hence the interpretation remains a subject for discussions \cite{Chang:2019ars,Leane:2019xiy,Leane:2020pfc,Dinsmore:2021nip,Ramirez:2024oiw,Hu:2025thq,Manconi:2025ogr,Holst:2024fvb}.

Recently, Totani \cite{Totani:2025fxx} studied gamma rays from the Milky Way halo based on the 15-year data from the Fermi satellite, painstakingly subtracting astrophysical sources of gamma rays. By avoiding the galactic disk and center, a potential contamination of gamma rays from astrophysical sources was minimized. He found statistically significant (5--8$\sigma$) flux of gamma rays whose morphology was consistent with the density squared in the halo. The energy spectrum is peaked around 20~GeV, consistent with annihilation of particles in the 500--800 GeV mass range, unlike falling spectrum expected from astrophysical sources. 

There are two puzzling issues in the claimed signal. In order to account for the gamma ray flux, the annihilation cross section in the halo was estimated to be
\begin{align}
	\langle \sigma_{ann} v_{rel} \rangle_{\rm MW}
	= (5\mbox{--}8) \times 10^{-25}~{\rm cm}^3/{\rm s}, \label{eq:sigmaMW}
\end{align}
more than an order of magnitude larger than \eqref{eq:sigmafo}. In addition, it is in tension with the upper limit from dwarf galaxies \cite{Fermi-LAT:2015att,Fermi-LAT:2016uux,MAGIC:2020ceg,McDaniel:2023bju}. 

Normally for a WIMP, a larger annihilation cross section implies it comprises only a fraction of dark matter abundance $f < 1$, and is acceptable if there are other components in the dark matter density. However it is not acceptable in this case because the required cross section needs to be boosted as $1/f^2$ to account for the observed flux. If we take the signal at its face value, it calls for an alternative interpretation.

There is a similar issue in self-interacting dark matter (SIDM) \cite{Spergel:1999mh} to explain discrepancy between the observed diverse density profiles and the N-body simulations \cite{Navarro:1996gj}. The limits from clusters of galaxies are stronger than what would be required for this purpose (see, {\it e.g.}\/, \cite{Tulin:2017ara} for a review). The current author proposed a resonant self-interaction to explain the velocity dependence in the dark matter self-interaction cross section with various collaborators \cite{Chu:2018fzy,Tsai:2020vpi,Kondo:2022lgg}. This work builds on this proposal for annihilation cross section rather than the self-scattering. A resonant enhancement in annihilation cross section was also explored in the literature by many authors \cite{Griest:1990kh,Gondolo:1990dk,Jungman:1995df,Feldman:2008xs,Pospelov:2008jd,Ibe:2008ye,March-Russell:2008klu,Guo:2009aj,Ibe:2009dx,Kakizaki:2005en}. 

In this letter, we explore the possibility to explain the discrepancy among annihilation cross sections in different environments because of a velocity dependence in the cross section due to a resonance. We perform a simple and model-independent analysis of this possibility. We also briefly touch on a simple toy particle-physics model that can accommodate required parameters. 

\begin{figure}[t]
\includegraphics[width=\columnwidth]{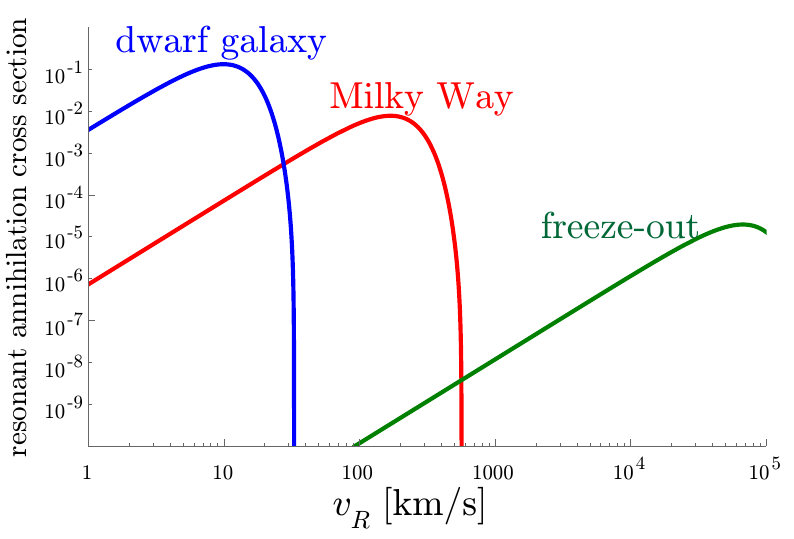}
\caption{The schematics of a resonant annihilation cross section in the freeze-out, Milky Way halo, and a typical dwarf galaxy, as a function of the resonant velocity $v_R$ (km/s). The vertical axis is in the unit of (km/s)$^{-1}$ and should be multiplied by $\sigma_R \frac{2\Gamma}{m}$ to obtain $\langle \sigma v_{rel} \rangle$. See Eq.~\eqref{eq:sigmaaverage2}. We took $v_0=10$~km/s and $v_{esc}=33$~km/s for a dwarf galaxy, $v_0=170$~km/s and $v_{esc}=567$~km/s for the Milky Way, and $x_f=20$ for the freeze-out. For $v_R$ in the range from a few tens to a few hundreds of km/s, the resonant annihilation is absent in dwarf galaxies, of little importance at the freeze-out, but is significant in the Milky Way halo.}
\label{fig:vRdep}
\end{figure}

\section{Resonant Annihilation}

A resonance in scattering of particles is an unstable state that is temporarily produced in the collision that decays into other particles. Vast majority of microscopic particles are resonances because of their finite lifetimes. 

The standard Breit--Wigner formula for a resonance is (see, {\it e.g.}\/, \cite{ParticleDataGroup:2024cfk}),
\begin{align}
	\sigma &= \frac{2J+1}{(2S_1+1)(2S_2+1)} \frac{4\pi\hbar^2}{k^2}
	\frac{\Gamma^2/4}{(E-E_R)^2 + \Gamma^2/4} B_{in} B_{out} ,
\end{align}
where $k$ is the momentum of each incoming particle and $E$ the total energy in the center-of-momentum (CM) frame.  $E_R = M c^2$ is the mass of the resonance, $J$ is the spin of the resonance, and $S_{1,2}$ are spins of incoming (dark matter) particles. $B_{in}$ is the branching fraction of the resonance decaying into the initial state (dark matter), and $B_{out}$ that into the final state (standard model). $\Gamma$ is the Full Width at Half Maximum (FWHM) in the energy dependence, and is related to the lifetime of the resonance as $\tau = \hbar/\Gamma$. This formula is not meant to be true at all momenta, but is rather a description in the vicinity of the resonance $E \approx E_R$.

For a narrow resonance, we can use the narrow width approximation,
\begin{align}
	\sigma &\approx \sigma_0 + 
	\sigma_R 	\frac{\pi \Gamma}{2} \delta (E- E_R) .
\end{align}
Here, $\sigma_0$ is the piece for the standard $S$-wave thermal freeze-out \eqref{eq:sigmafo}, while
\begin{align}
	\sigma_R &= \frac{(2J+1)B_{in} B_{out} }{(2S_1+1)(2S_2+1)} \frac{4\pi\hbar^2}{k^2}.
\end{align}

We are interested in a resonance for non-relativistic incoming particles, and hence
\begin{align}
	E &= 2\left(mc^2 + \frac{1}{2}m v_R^2 \right) = E_R = Mc^2
\end{align}
in the CM frame. Therefore, the resonant velocity is
\begin{align}
	v_R^2 = c^2 \frac{M - 2m}{m}\ .
\end{align}
We study the regime where $M-2m \ll m$ so that $v_R \ll c$. 

The main idea for a resonant annihilation is depicted schematically in Fig.~\ref{fig:vRdep}. Given the large differences among the velocity distributions of the freeze-out process, Milky Way halo, and halos of dwarf galaxies, a resonant annihilation cross section peaked at a particular velocity would lead to very different averaged annihilation cross sections in three enviornments. %In this plot we used the parameters extracted later in this letter Eqs.~(\ref{eq:vR},\ref{eq:GammaBB2}).

\section{Model-Independent Analysis}

We need to compute the averaged cross section due to the resonant piece. Both in the thermal bath and the halo, we approximate the velocity distribution to be Maxwellian,
\begin{align}
	p(v) d^3 v = \frac{1}{(2\pi)^{3/2} v_0^3} e^{-v^2/2v_0^2} d^3 v.
\end{align}
For incoming particles of velocities $\vec{v}_{1,2}$, we can define
\begin{align}
	\vec{V} = \frac{1}{2} (\vec{v}_1 + \vec{v}_2), \qquad
	\vec{v}_{rel} = \vec{v}_1 - \vec{v}_2.
\end{align}
The integration volume is $d^3 v_1 d^3 v_2 = d^3 V d^3 v_{rel}$. We find
\begin{align}
	p(v_1) p(v_2) d^3 v_1 d^3 v_2
%	&= \frac{1}{(2\pi)^3 v_0^6} 
%	e^{-(\vec{V}+\frac{1}{2}\vec{v}_{rel})^2/2v_0^2} e^{-(\vec{V}-\frac{1}{2}\vec{v}_{rel})^2/2v_0^2}
%	d^3 V d^3 v_{rel} \nonumber \\
	&=  \frac{1}{(2\pi)^3 v_0^6} 
	e^{-\vec{V}^2/v_0^2} e^{-\vec{v}_{rel}^2/4v_0^2}
	d^3 V d^3 v_{rel} \nonumber \\
%	&= \frac{1}{(4\pi)^{3/2} v_0^3}  e^{-\vec{v}_{rel}^2/4v_0^2} d^3 v_{rel} \nonumber \\
	&= \frac{1}{\pi^{3/2} v_0^3}  e^{-\vec{v}^2/v_0^2} d^3 v,
\end{align}
where $\vec{v}=\pm \frac{1}{2}\vec{v}_{rel}$ are the velocities of two particles in the CM frame and the center-of-mass velocity $\vec{V}$ is integrated over. Therefore,
\begin{align}
	\langle \sigma v_{rel} \rangle_R
	&=  \int  \sigma_R
	\frac{\Gamma^2/4}{(E-E_R)^2 + \Gamma^2/4} 2 v
	\frac{1}{\pi^{3/2} v_0^3}  e^{-\vec{v}^2/v_0^2} d^3 v \nonumber \\
	&\approx  \int \sigma_R
	\frac{\pi\Gamma}{2} \delta\left( m (v^2 - v_R^2) \right) 2v
	\frac{1}{\pi^{3/2} v_0^3}  e^{-\vec{v}^2/v_0^2} 4\pi v^2 d v \nonumber \\
%	&=  \frac{(2J+1)B_{in} B_{out}}{(2S_1+1)(2S_2+1)} \frac{4\pi\hbar^2}{(mv_R)^2}
%	\frac{\pi\Gamma}{2} \frac{1}{2mv_R} 
%	\frac{1}{\pi^{3/2} v_0^3}  e^{-v_R^2/v_0^2} 2 v_R 4\pi v_R^2 \nonumber \\
	&=  \frac{2\pi^{1/2} v_R^2\Gamma}{m v_0^3}
	e^{-v_R^2/v_0^2}  \sigma_R.
	\label{eq:sigmavrelR}
\end{align}
When used for a thermal average at the time of the freeze-out, the Boltzmann factor is $e^{-m \vec{v}^2/2kT}$ and hence $v_0^2 =\frac{kT_f}{m} = c^2 x_f^{-1}$, where typically $x_f \approx 20$. 
%\end{widetext}

\begin{widetext}
On the other hand, the expression \eqref{eq:sigmavrelR} is not appropriate if the escape velocity $v_{esc}$ is comparable or below the resonant velocity in a galactic halo, 
\begin{align}
	\langle \sigma v_{rel} \rangle_R
	&\approx  \int_0^{v_{esc}}  \sigma_R 
	\frac{\pi\Gamma}{2} \delta\left( m (\frac{1}{4}(\vec{v}_1-\vec{v}_2)^2 - v_R^2) \right) 2 v
	\frac{1}{(2\pi)^3 v_0^6}  e^{-\vec{v}_1^2/2v_0^2} e^{-\vec{v}_2^2/2v_0^2} 
	d^3 v_1 d^3 v_2 \nonumber \\
	&=  \sigma_R
	\frac{\pi\Gamma}{2} 2v_R  \frac{1}{(2\pi)^3 v_0^6} 
	\int_0^{v_{esc}}  
	\delta\left( \frac{1}{4} m (v_1^2+ v_2^2 -2v_1 v_2 \cos\theta - 4v_R^2) \right)  
	e^{-\vec{v}_1^2/2v_0^2} e^{-\vec{v}_2^2/2v_0^2} 
	4\pi v_1^2 d v_1 2\pi v_2^2 d v_2 d\cos\theta .
\end{align}
%\end{widetext}
The $\cos\theta$ integral can hit the delta function only if
\begin{align}
	|v_1-v_2| \leq 2 v_R \leq v_1+v_2.
\end{align}
After carrying out the rest of the integral within this bound, we obtain
%\begin{widetext}
\begin{align}
	\langle \sigma v_{rel} \rangle
	&=  \sigma_R \frac{\Gamma}{v_0^3} \frac{2v_R}{m}  
	\left( e^{-v_{esc}^2/v_0^2} \left( 1 - e^{2(v_{esc}-v_R)v_R/v_0^2}\right) v_0
	+ e^{-v_R^2/v_0^2} \sqrt{\pi}\  {\rm erf}\left( \frac{v_{esc}-v_R}{v_0} \right) v_R \right).
	\label{eq:sigmaaverage2}
\end{align}
It is easy to verify that this expression reduces to \eqref{eq:sigmavrelR} in the limit $v_{esc} \rightarrow \infty$. On the other hand, it goes to zero when $v_{esc} \rightarrow v_R$. 
\end{widetext}

We plot the averaged resonant cross section $\langle \sigma v_{rel} \rangle_R$ for the freeze-out, Milky Way halo, and a typical dwarf galaxy halo in Fig.~\ref{fig:vRdep}, leaving a factor $\sigma_R 2\Gamma/m$ out. Here we took the parameters $v_0=10$~km/s and $v_{esc}=33$~km/s for a dwarf galaxy, $v_0=170$~km/s and $v_{esc}=567$~km/s for the Milky Way, and $x_f=20$ for the freeze-out. There is a wide range of $v_R$ where the resonant contribution is absent in dwarf galaxies, little at the freeze-out, but is important in the Milky Way halo. We will take $v_R=100$~km/s as a canonical choice below for numerical evaluations. 

If the resonance velocity is above the escape velocities in dwarf galaxies of a few tens of km/s, the average annihilation cross section in dwarf galaxies reduces back to the velocity-independent piece \eqref{eq:sigmafo} without the resonant contribution. It is compatible with the upper limit from dwarf galaxies \cite{Fermi-LAT:2015att,Fermi-LAT:2016uux,MAGIC:2020ceg,McDaniel:2023bju} for the mass range 500--800~GeV reported in \cite{Totani:2025fxx}. For $v_R=100$~km/s, the annihilation in the Milky Way halo Eq.~\eqref{eq:sigmaMW} requires
\begin{align}
	\sigma_R &= 1.05 \times 10^{-38}~{\rm cm}^2 
	\frac{mc^2}{\rm 0.8~TeV} \frac{\rm TeV}{\Gamma} 
	\frac{\langle \sigma v_{rel}\rangle_{\rm MW}}{6 \times 10^{-25}~{\rm cm}^3{\rm s}^{-1}}\ .
	\label{eq:sigmaR}
\end{align}

%The main point in this letter is that the enhanced annihilation cross section in the Milky Way halo \eqref{eq:sigmaMW} can be explained by the resonant term in Eq.~\eqref{eq:sigmaaverage}. Using $\sigma_{\rm MW} \approx 170$~km/s  \cite{Folsom:2025lly} (note the difference in notation $\sigma_{MW} = \sqrt{2}\sigma$), we need
%\begin{align}
%	v_R^2\Gamma\sigma_R
%	= \frac{\sigma_{MW}^3}{2\pi^{1/2}}  m \langle \sigma_{ann} v_{rel} \rangle_{\rm MW} .
%	\label{eq:MWrequirement}
%\end{align}

%
%
%In the thermal bath, the Boltzmann factor is $e^{-m \vec{v}^2/2kT}$ and hence $\sigma^2 =\frac{kT}{m}$. In the early universe when $v_R \ll \frac{kT}{m}$, the thermally averaged cross section can be rewritten using the requirement \eqref{eq:MWrequirement},
%\begin{align}
%	\langle \sigma v_{rel} \rangle 
%	& = \langle \sigma_0 v_{rel} \rangle +2\pi^{1/2} v_R^2\Gamma \left( \frac{m}{(kT)^{3}}\right)^{1/2}
%	\sigma_R \nonumber \\
%	&= \langle \sigma_0 v_{rel} \rangle + \left( \frac{m}{kT}\right)^{3/2}
%	\sigma_{MW}^3 \langle \sigma_{ann} v_{rel} \rangle_{\rm MW}.
%\end{align} 

\begin{figure}[t]
\includegraphics[width=\columnwidth]{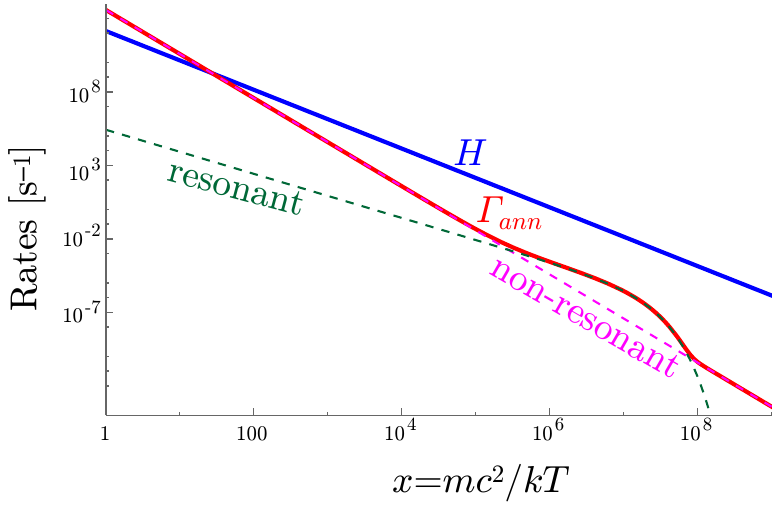}
\caption{The expansion rate $H$ of the radiation dominated universe and the annihilation rate $\Gamma_{ann}$ as a function of $x=mc^2/kT$, where $m$ is the mass of the dark matter.  Later times correspond to larger $x$ and hence to the right of the plot. The annihilation rate has a temporary bump due to the resonance which disappears after the temperature drops below the resonance energy. Dashed lines show the breakdown of the annihilation rate into the resonant and non-resonant pieces.}
\label{fig:rates}
\end{figure}

One may be concerned that the resonant piece becomes important long after the freeze-out and dark matter may start annihilating again. However, this is not the case. We show the annihilation rate and the expansion rate as a function of the temperature in Fig.~\ref{fig:rates}. The resonant cross section becomes more important for $x\simeq 10^5$--$10^8$ but before it becomes as fast as the expansion rate, it dies down because of the Boltzmann suppression to hit the resonance energy. 

To reproduce the required $\sigma_R$ \eqref{eq:sigmaR}, we need
\begin{align}
	\Gamma B_{in} B_{out}
	&= 1.53 \times 10^{-10}~{\rm GeV} \left( \frac{mc^2}{\rm 800~GeV}\right)^3
	\nonumber \\
	& \times \left(\frac{\rm 100~km/s}{v_R}\right)^2
	\frac{\langle \sigma v_{rel} \rangle_{\rm MW}}{6\times 10^{-25}~{\rm cm}^3{\rm s}^{-1}}\ .
\end{align}

%There are several constraints on these parameters. One of them is that the resonant annihilation in the early universe goes down slowly as $(kT)^{3/2}$ while the expansion rate faster as $(kT)^2$. 

%\begin{widetext}
%This is appropriate for the time of the freeze-out, $x_f = mc^2/kT \approx 20$ and hence $\sigma_{\rm f.o.} = c x_f^{-1/2} \approx 67,000$~km/s. To reproduce \eqref{eq:sigmafo}, we find
%\begin{align}
%	\frac{(2J+1)B_{in} B_{out}}{(2S_1+1)(2S_2+1)} \Gamma 
%	&= 2.42 \times 10^{-7}~{\rm TeV} 
%	\left( \frac{m}{0.8~{\rm TeV}} \right)^3 . \label{eq:Gamma}
%\end{align}
%Here we used the fact that $e^{v_R^2/\sigma_{\rm f.o.}^2}\approx 1$ because we will find $v_R \ll \sigma_{\rm f.o.}$ later. 
%
%
%
%What we need in the Milky Way halo is Eq.~\eqref{eq:sigmaMW} with $\sigma_{\rm MW} \approx 170$~km/s with the escape velocity $v_{esc} \approx 567$~km/s \cite{Folsom:2025lly} (note the difference in notation $v_0 = \sqrt{2}\sigma$). Using the result \eqref{eq:Gamma} and to reproduce the reported annihilation cross section in the halo \eqref{eq:sigmaMW}, we find
%\begin{align}
%	v_R &= 552\mbox{--}555~\mbox{km/s}.
%	\label{eq:vR}
%\end{align}
%This resonance velocity is well above the escape velocities in dwarf galaxies of a few tens of km/s, and hence the average annihilation cross section in dwarf galaxies is zero within the narrow width approximation. Therefore the existing upper limits are trivially satisfied. 
%

%
%It is easy to understand why the preferred resonant velocity is close to the escape velocity. The annihilation cross sections (\ref{eq:sigmaaverage},\ref{eq:sigmaaverage2}) are proportional to $1/\sigma^3$ and hence the inferred value for the Milky Way halo from the freeze-out cross section tends to be larger by a factor of $(\sigma_{\rm f.o.}/\sigma_{\rm MW})^3 \approx 6.1 \times 10^{7}$, see Fig.~\ref{fig:sigmaMW}, which is an overkill. By placing the resonant velocity close to the maximum given by the escape velocity, the enhancement is not as prominent as it could be. This is due to our assumption that the annihilation cross section is purely the resonant piece. In many models, the resonant piece is often accompanied by non-resonant pieces with no or milder velocity dependences. If the velocity-independent piece (dominant) comes with a resonant piece (sub-dominant), the resonant velocity does not have to be this close to the escape velocity. Our study should be viewed as a demonstration that the level of enhancement reported in \cite{Totani:2025fxx} can be easily accommodated by a resonant piece in the annihilation process.

\section{A Model}

Here we conduct a sanity test that the parameters determined in the above model-independent analysis can be accommodated in a particle physics model. 

The dark matter is a scalar field $\chi$ with the potential
\begin{align}
	V_\chi = \frac{1}{2} m^2 \chi^2 + \frac{\lambda}{2} \chi^2 H^\dagger H .
\end{align}
There is a ${\mathbb Z}_2$ symmetry $\chi \rightarrow -\chi$ which makes $\chi$ stable and hence the dark matter. $H$ is the Higgs doublet in the standard model. The $\chi$ annihilates dominantly to $\chi\chi \rightarrow h h, W^+ W^-, ZZ$ and the annihilation cross section can be worked out using the equivalence theorem limit $m \gg m_{W,Z}$,
\begin{align}
	\sigma v_{rel} & =  \frac{\lambda^2}{16\pi m^2} \ .
\end{align}
To reproduce Eq~\eqref{eq:sigmafo}, we need
\begin{align}
	\lambda = 0.25 \frac{mc^2}{\rm 800~GeV}\ .
\end{align}
The fact that the required annihilation cross section can be obtained without an extreme choice of parameters is exactly the attractive nature of the WIMP hypothesis. 

To have a resonant annihilation, we introduce a weakly coupled scalar field $\Sigma$ of mass $M c^2 = 2 m c^2 + m v_R^2$ with an additional potential
\begin{align}
	V_\Sigma &= \frac{1}{2} M^2 \Sigma^2 + \frac{1}{2} \mu \Sigma \chi^2  .
\end{align}
If we also assume a small mixing $\epsilon$ between $\Sigma$ and the standard model Higgs boson, the annihilation can proceed through the $\Sigma$ resonance. For simplicity, we assume $B_{in} =  B_{out} = \frac{1}{2}$, and $S_1=S_2=J=0$. Then we need 
\begin{align}
	\Gamma B_{in} = \Gamma B_{out} = 3.1 \times 10^{-13}~{\rm TeV} .
	\label{eq:GammaBB2}
\end{align}
The decay width for the initial state is
\begin{align}
	\Gamma B_{in} &= \Gamma(\Sigma \rightarrow \chi \chi) = \frac{1}{2!} \frac{1}{2M} \mu^2 \frac{\beta}{8\pi}
%	\nonumber \\
	= \frac{\mu^2}{32\pi M} \frac{v_R}{c} .
\end{align}
which requires $\mu = 0.38$~GeV. On the other hand, $\Sigma$ can decay $ \Sigma \rightarrow h^* \rightarrow W^+ W^-, ZZ$, where $h^*$ is an off-shell standard model Higgs boson,
\begin{align}
	\Gamma B_{out}
	&=\Gamma(\Sigma \rightarrow h^* \rightarrow W^+ W^-, Z Z) 
	\nonumber \\
	& = \epsilon^2 \frac{3}{16\sqrt{2} \pi} G_F M^3 
	= \epsilon^2 2.0 {\rm TeV} \left( \frac{M}{1.6~{\rm TeV}} \right)^3,
\end{align}
and $\epsilon = 3.9 \times 10^{-7}$ is required. Both $\mu$ and $\epsilon$ are small, which is technically natural given that both of them violate ${\mathbb Z}_2$ symmetry $\Sigma \rightarrow -\Sigma$. Note also that the final states are very similar to $b\bar{b}$ and $W^+W^-$ considered by Totani because the Higgs boson overwhelmingly decays into $b\bar{b}$ and the $Z$ branching fractions are similar to those of the $W^\pm$ boson. 
%The width 
%\begin{align}
%\Gamma(\sigma \rightarrow \pi \pi) &\sim \frac{\lambda^2}{8\pi} m_\sigma c^2 \frac{v_R}{c}
%= 0.022~{\rm TeV},
%\end{align}
%where $\beta = v_R/c \approx 0.002$ and a strong coupling $\lambda \approx 4\pi$ is assumed. 

This level of mixing $\epsilon$ is perfectly consistent with the current level of precision on Higgs boson couplings at the LHC \cite{HaiderAbidi2025}. If we relax the assumption $B_{in}=B_{out} = \frac{1}{2}$ and accept larger $B_{out} = B(\Sigma \rightarrow W^+ W^-, Z Z)$, the mixing angle can be larger and hence a target for more precise measurements in the Higgs boson couplings at HL-LHC and future proposed $e^+ e^-$ Higgs factories such as FCC-ee, CEPC, or linear colliders. 

\section{Discussion}

The nearly exact factor of two between the dark matter and resonance $M-2m \ll m$ may appear fine-tuned. There are various ways such a factor may appear. 

The most intuitive way is that the dark matter and the resonance are made of the same constituents \cite{Tsai:2020vpi}. Examples in hadronic physics include $\phi \rightarrow K^+ K^-$, $D^* \rightarrow D \pi$, $B_{s1} \rightarrow B^* K^0$, $\Upsilon(4S) \rightarrow B^0 \bar{B}^0$, some at a permille level accuracy. In nuclear physics there are more accurate cases such as $^8{\rm Be} \rightarrow 2\alpha$ with the precision of $0.000012$. 

Another possibility is that both the dark matter and the resonance are Kaluza--Klein states of flat extra dimensions \cite{Chu:2018fzy}. If the extra dimension is a circle $S^1$ of radius $R$, the Kaluza--Klein states have masses $\frac{2\pi}{R} n$ where $n$ is an integer. The states of $n=1$ and $n=2$ have their masses with a ratio of two. Radiative corrections shift their masses by a small amount while maintaining the near threshold spectrum \cite{Lee:2025lko}. 

Some masses can be accidentally resonant. For instance, the $\sigma$ state (now renamed $f_0(500)$ by Particle Data Group \cite{ParticleDataGroup:2024cfk}) in the light hadron spectrum appears as a resonance in the $\pi\pi$ scattering. Its existence has been controversial for half a century but is now established both experimentally \cite{Pelaez:2015qba} and theoretically by lattice simulations \cite{Briceno:2016mjc,Briceno:2017qmb}. It is believed to be a tetraquark $(ud)(\bar{u}\bar{d})$ \cite{Jaffe:1976ig} or meson molecule $(\pi\pi)$ state. As the quark mass is increased, $m_\pi^2$ grows linearly with quark mass, while $m_\sigma^2$ decreases linearly with quark mass, crossing the threshold $m_\sigma = 2 m_\pi$ \cite{Kondo:2022lgg,Kondo:2025njf}. This can also be used for a resonant annihilation.  

We also add that the Sommerfeld enhancement \cite{Hisano:2003ec} may be a good way to reconcile Milky Way halo and freeze-out cross sections. It would require a separate analysis and is beyond the scope of this letter.

\section{Conclusion}

In this letter, we explored a possible explanation to the discrepancy in WIMP annihilation cross sections between the freeze-out, Milky Way halo, and dwarf galaxies. We find that a resonant annihilation can reconcile them when the annihilation proceeds through a resonance within the narrow width approximation, in addition to a standard velocity-independent piece. We mentioned a possible test at collider experiments.

%\section*{\textbf{Acknowledgements}}

\acknowledgments

This work was supported by the NSF grant PHY-2515115, by the Director, Office of Science, Office of High Energy Physics of the U.S. Department of Energy under the Contract No. DE-AC02-05CH11231, by the JSPS Grant-in-Aid for Scientific Research JP23K03382, Hamamatsu Photonics, K.K, Tokyo Dome Corporation, and by the World Premier International Research Center Initiative (WPI) MEXT, Japan.

\bibliography{refs}

\end{document}